\documentclass[twocolumn]{emulateapj}
\usepackage{apjfonts}

\usepackage{natbib}
\usepackage{epsfig}

\newcommand{\be}{\begin{equation}}
\newcommand{\ee}{\end{equation}}

\newcommand{\eg}{\emph{e.g.}}
\newcommand{\kms}{\mbox{km\ \ensuremath{\rm{s}^{-1}}}}

\shortauthors{Cordiner et al.}

\begin{document}

\title{ALMA measurements of the HNC and HC$_3$N distributions in Titan's atmosphere}

\submitted{ApJ Letters 2014, in press}

\author{M. A. Cordiner\altaffilmark{1,2}, C. A. Nixon\altaffilmark{1}, N. A. Teanby\altaffilmark{3}, P. G. J. Irwin\altaffilmark{4}, J. Serigano\altaffilmark{1}, S. B. Charnley\altaffilmark{1}, S. N. Milam\altaffilmark{1}, M. J. Mumma\altaffilmark{1},  D. C. Lis\altaffilmark{5,6},  G. Villanueva\altaffilmark{1}, L. Paganini\altaffilmark{1}, Y.-J. Kuan\altaffilmark{7,8}, A. J. Remijan\altaffilmark{9}}

\altaffiltext{1}{NASA Goddard Space Flight Center, 8800 Greenbelt Road, Greenbelt, MD 20771, USA.}
\email{martin.cordiner@nasa.gov}
\altaffiltext{2}{Department of Physics, Catholic University of America, Washington, DC 20064, USA.}
\altaffiltext{3}{School of Earth Sciences, University of Bristol, Wills Memorial Building, Queen's Road, Bristol, BS8 1RJ, UK.}
\altaffiltext{4}{Atmospheric, Oceanic, and Planetary Physics, Clarendon Laboratory, University of Oxford, Parks Road, Oxford, OX1 3PU, UK}
\altaffiltext{5}{Sorbonne Universit{\'e}s, Universit{\'e} Pierre et Marie Curie, Paris 6, CNRS, Observatoire de Paris, UMR 8112, LERMA, Paris, France.}
\altaffiltext{6}{California Institute of Technology, Cahill Center for Astronomy and Astrophysics 301-17, Pasadena, CA~91125, USA}
\altaffiltext{7}{National Taiwan Normal University, Taipei 116, Taiwan, ROC.}
\altaffiltext{8}{Institute of Astronomy and Astrophysics, Academia Sinica, Taipei 106, Taiwan, ROC.}
\altaffiltext{9}{National Radio Astronomy Observatory, Charlottesville, VA 22903, USA.}

\begin{abstract}

We present spectrally and spatially-resolved maps of HNC and HC$_3$N emission from Titan's atmosphere, obtained {using} the Atacama Large Millimeter/submillimeter Array (ALMA) on 2013 November 17.  These maps show anisotropic spatial distributions for both molecules, with resolved emission peaks in Titan's northern and southern hemispheres. The HC$_3$N maps indicate enhanced concentrations of this molecule over the poles, consistent with previous studies of Titan's photochemistry and atmospheric circulation. Differences between the {spectrally-}integrated flux distributions of HNC and HC$_3$N show that these species are not co-spatial. The observed spectral line shapes are consistent with HNC {being concentrated} predominantly in the mesosphere and above (at altitudes $z\gtrsim 400$~km), whereas HC$_3$N is abundant at a broader range of altitudes ($z\approx70$-600~km). From spatial variations in the HC$_3$N line profile, the locations of the HC$_3$N emission peaks are shown to be variable as a function of altitude.  {The peaks in the integrated emission from HNC and the line core (upper-atmosphere) component of HC$_3$N} (at $z\gtrsim300$~km) are found to be asymmetric with respect to Titan's polar axis, indicating that the mesosphere may be more longitudinally-variable than previously thought. {The spatially-integrated HNC and HC$_3$N} spectra are modeled using the {NEMESIS} planetary atmosphere code and the resulting best-fitting disk-averaged vertical mixing ratio (VMR) profiles are found to be in reasonable agreement with previous measurements for these species. {Vertical column densities of the best-fitting gradient models for HNC and HC$_3$N are $1.9\times10^{13}$~cm$^{-2}$ and $2.3\times10^{14}$~cm$^{-2}$, respectively.}

\end{abstract}


\section{Introduction}

Previous remote and in-situ measurements have identified a wealth of molecules in Titan's atmosphere. Simple and complex hydrocarbons, nitriles and oxygen-bearing species have been detected in surprisingly large abundances, revealing the presence of a complex photo-chemistry that is considered as a possible analogue for the chemistry of the early Earth's atmosphere. Despite detailed studies of Titan's atmosphere by the Cassini spacecraft \citep[reviewed by][]{bez14}, major gaps in our understanding are still present regarding the full chemical inventory and the spatial distributions of Titan's chemical species. With its unprecedented combination of high spectral and spatial resolution at high sensitivity, the Atacama Large Millimeter/Submillimeter Array (ALMA) can obtain detailed measurements of the abundances and distributions of molecules inaccessible to Cassini and other ground and space-based instruments, as well as providing `instantaneous' maps (in a matter of minutes), of Titan's entire {Earth-facing} hemisphere.

Recent work on heterodyne millimeter-wave spectroscopy of Titan \citep{gur00,mar02,gur04,gur11,mor12} has proven the utility of high spectral resolution molecular line observations for the derivation of vertical mixing profiles, based on the variation of emission line shapes as a function of altitude. In the troposphere and stratosphere (at altitudes $z$ less than a few hundred kilometers), pressure broadening in the high-density gas (at N$_2$ number densities $n\gtrsim10^{15}$ cm$^{-3}$) gives rise to prominent line wings, whereas in the upper atmosphere ($z\gtrsim400$~km), densities are lower and rotational line profiles are typically dominated by the thermal and bulk motions of the gas, with narrow widths (FWHM $\lesssim1$~km/s).

The HNC molecule was detected in Titan's atmosphere for the first time by \citet{mor11}, who observed the $J=6-5$ rotational transition at 544~GHz using the Herschel Space Observatory. The telescope's angular resolution ($\approx39''$) was insufficient to spatially resolve the molecular emission, but line profile modeling indicated that the majority of the observed emission originated from HNC at altitudes greater than 400~km. Chemical models show that a likely pathway to HNC production in the ionosphere is from protonation of the abundant HCN molecule, followed by dissociative recombination of electrons with HCNH$^+$. This theory is supported by the detection of HCNH$^+$ at altitudes 1100-1150~km by the Cassini INMS instrument \citep{vui07}. {Alternatively, \citet{heb12} determined that the reaction H$_2$CN + H~$\longrightarrow$~HNC + H$_2$} could plausibly explain much of the observed HNC. Production of HNC from the degradation of large molecules and organic {polymers}/grains is also possible. {Measurement of the spatial distribution of} HNC is important in order to help constrain photochemical models and test theories regarding the formation and evolution of this molecule {on Titan}.

Infrared emission from HC$_3$N was detected on Titan by \citet{kun81}. The first radio measurement of this molecule was by \citet{bez92}, who observed the $J=16-15$ line at 146~GHz using the IRAM 30-m telescope. \citet{mar02} subsequently measured multiple (spectrally-resolved) transitions of HC$_3$N at 90-240~GHz and derived disk-averaged vertical mixing profiles for this molecule up to an altitude of 500~km. The major source of HC$_3$N on Titan is predicted to be from the reaction of the CN radical with C$_2$H$_2$ \citep{wil04,kra09}.

To date, the highest published spatial resolution achieved in ground-based observations of molecular emission from Titan's atmosphere was $0.8''\times0.4''$ with the IRAM Plateau de Bure interferometer \citep{mor05}. Variations in the HC$_3$N and CH$_3$CN line profiles over Titan's disk were observed, consistent with prograde equatorial wind velocities of $160\pm60$~m\,s$^{-1}$ and $60\pm20$~m\,s$^{-1}$ at altitudes of 300 km (upper stratosphere) and 450 km (lower mesosphere), respectively.

Here we present the first spectrally and spatially-resolved observations of sub-mm emission from the HNC and HC$_3$N molecules in Titan's atmosphere using ALMA, obtained with an angular resolution of approximately $0.5''\times0.3''$. Molecular emission maps and disk-averaged vertical mixing profiles are derived.

\begin{figure*}
\centering
\includegraphics[width=0.49\textwidth]{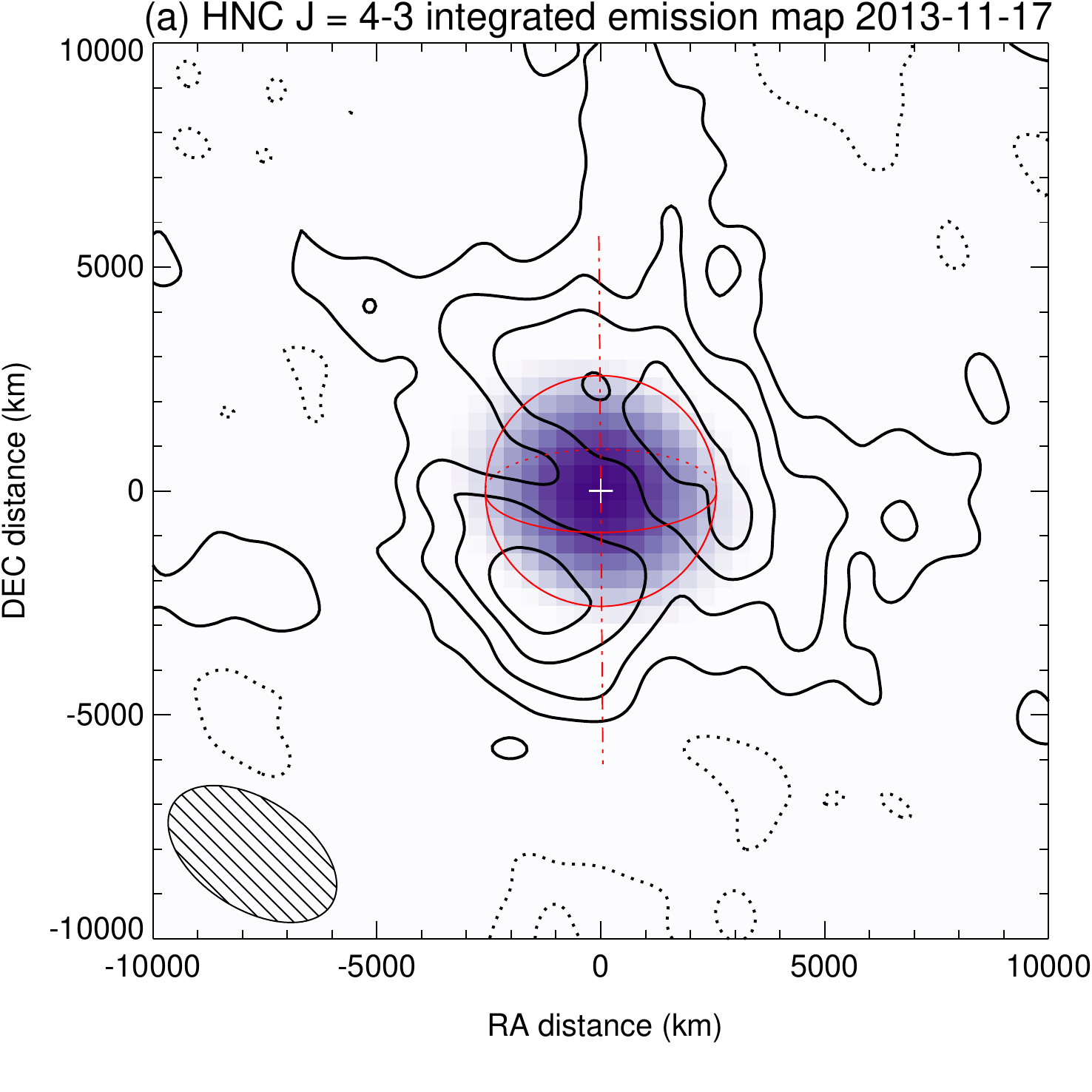}
\includegraphics[width=0.49\textwidth]{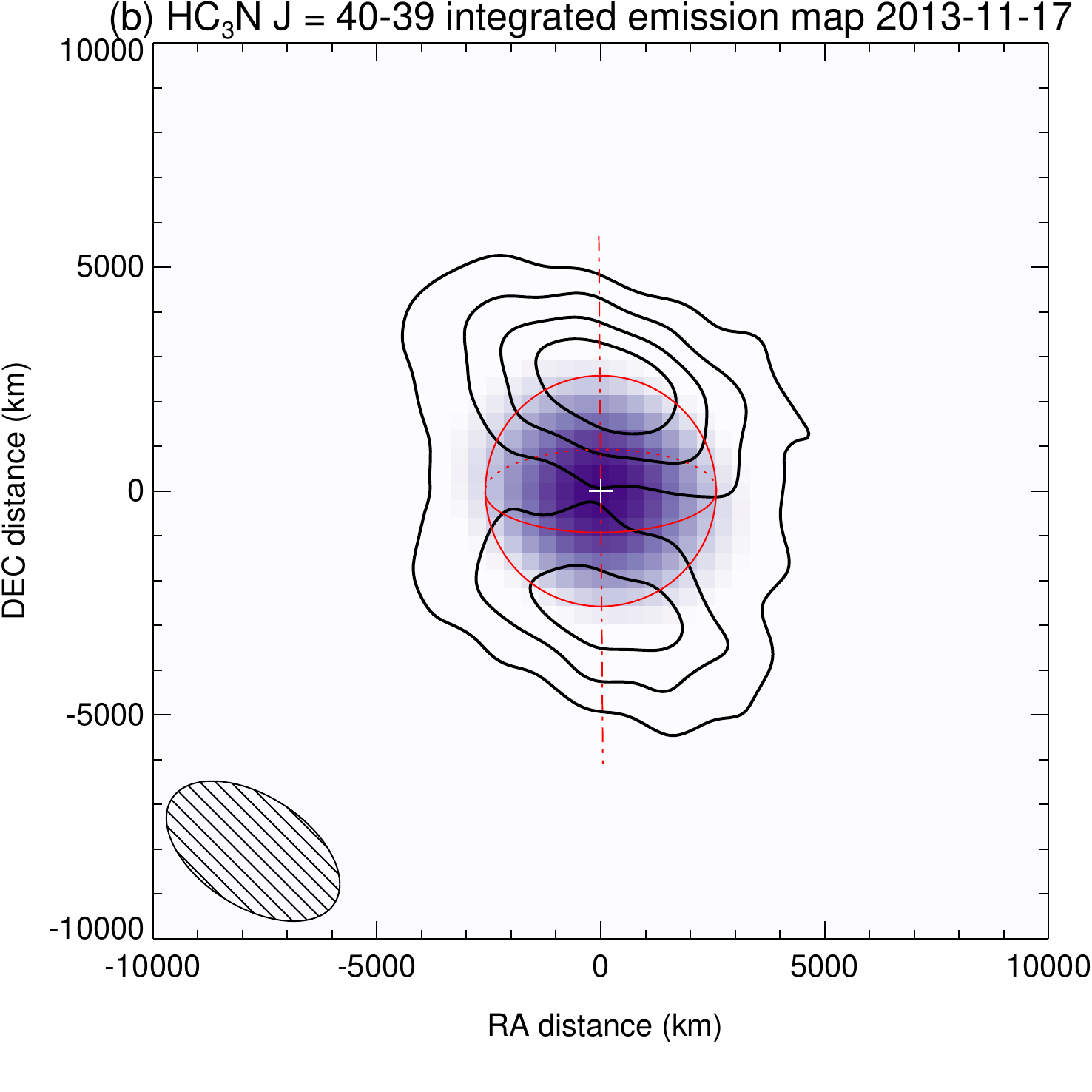}
\caption{(a) HNC $J=4-3$ and (b) HC$_3$N $J=40-39$ integrated flux maps of Titan (contours) observed on 2013-11-17, overlaid on the (simultaneously observed) 0.9~mm continuum image (purple bitmaps). The coordinate scale is in Titan-projected distances, with axes aligned with the RA-Dec grid. Contour levels are 20\%, 40\%, 60\% and 80\% of the peak intensity in each map. Contour separation is $2.1\sigma$ for HC$_3$N and $1.4\sigma$ for HNC, where $\sigma$ is the RMS noise (140~mJy\,\kms\,beam$^{-1}$ and 34~mJy\,\kms\,beam$^{-1}$, respectively). Negative contours are shown with dotted lines. The red circle represents Titan's disk (with equator marked), and the dot-dashed line is the polar axis, oriented 0.4$^{\circ}$ counter-clockwise from vertical, with the north pole tilted towards the observer by $21^{\circ}$. Titan's disk was almost fully illuminated at the time of observation, with a phase angle of $1.0^{\circ}$. The {PSF FWHM} ($0.54\times0.31''$), and orientation is shown lower left. \label{fig:maps}}
\end{figure*}

\begin{figure*}
\centering
\hspace{-2mm}
\includegraphics[width=0.45\textwidth]{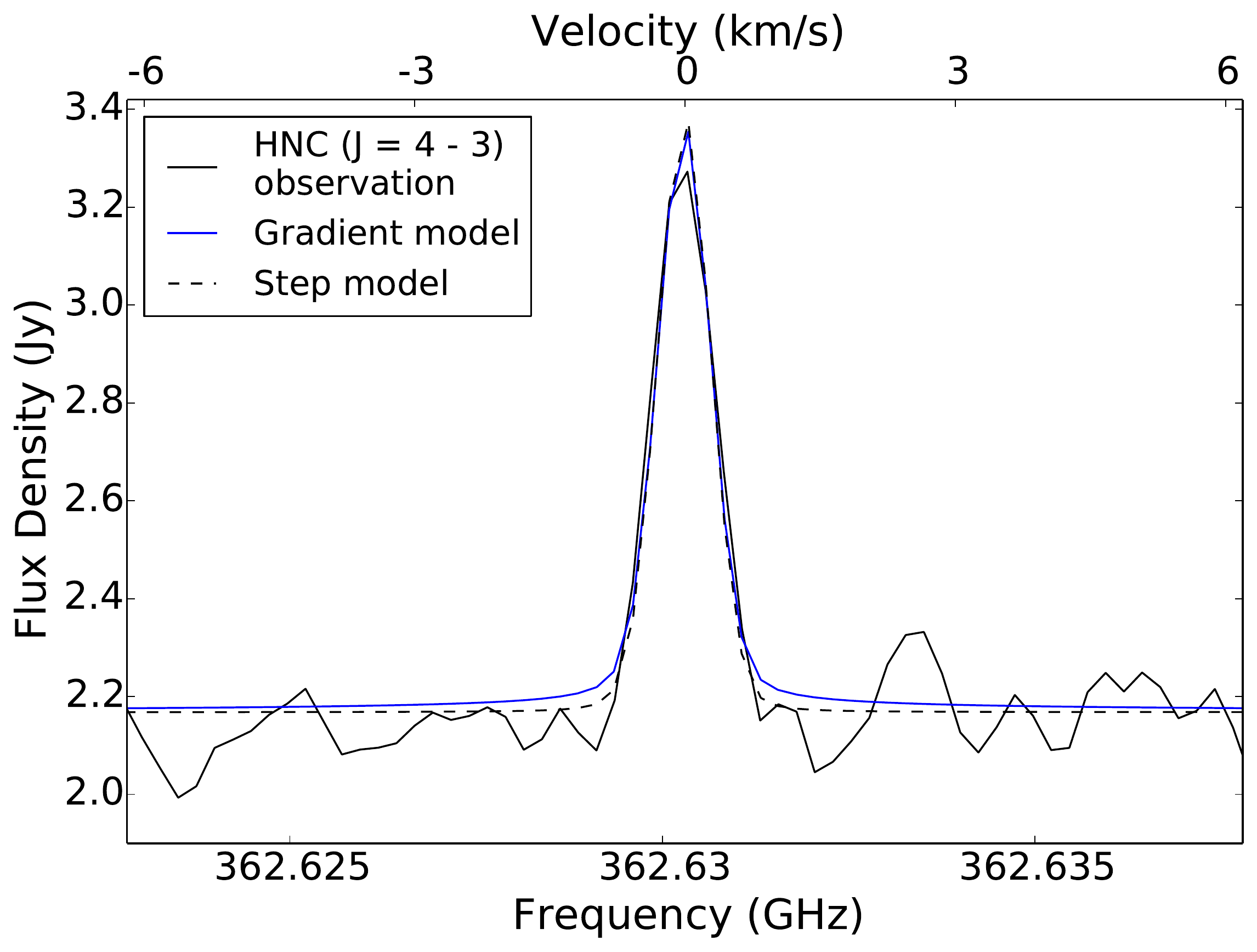}
\hspace{5mm}
\includegraphics[width=0.45\textwidth]{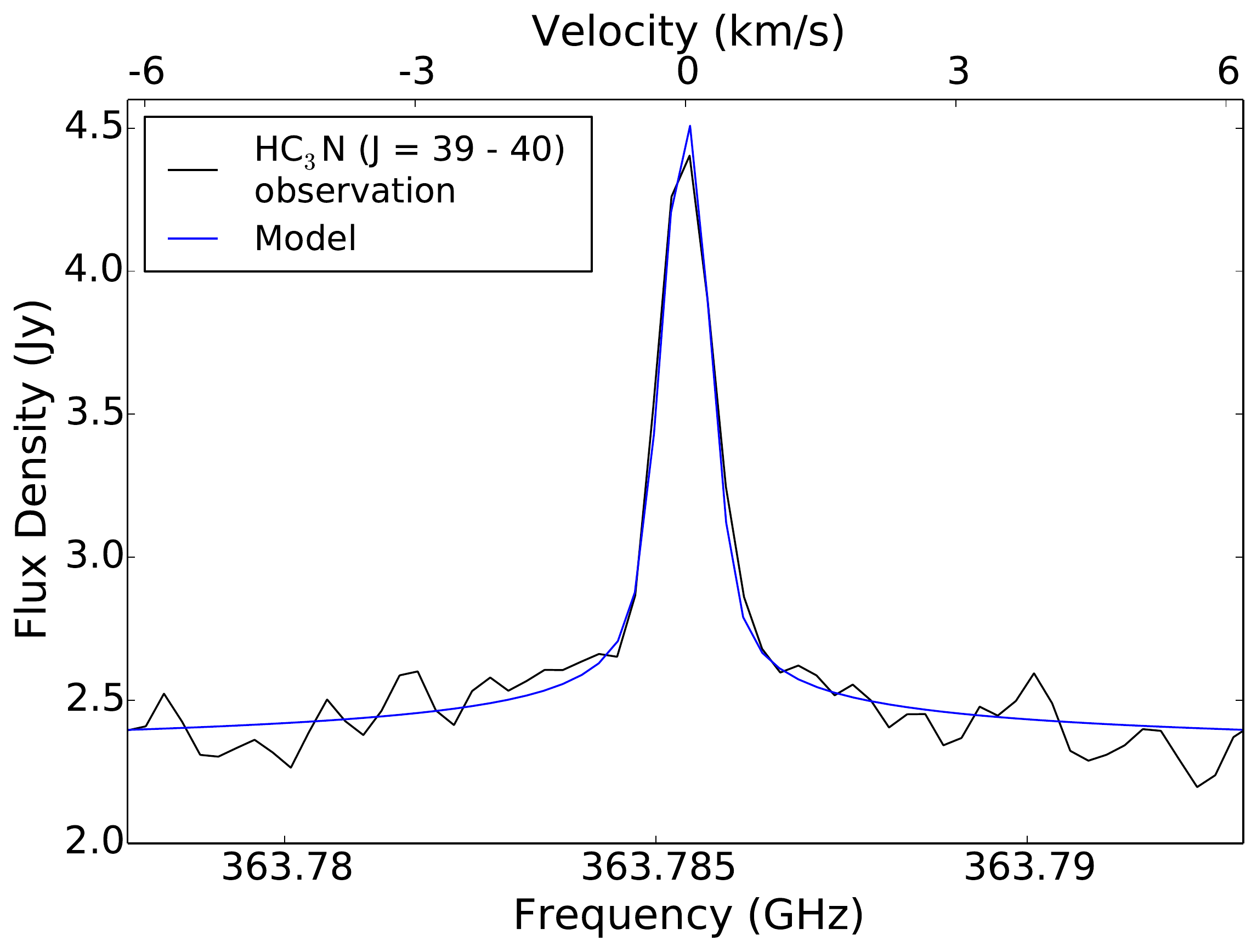}
\\
\vspace{5mm}
\includegraphics[width=0.47\textwidth]{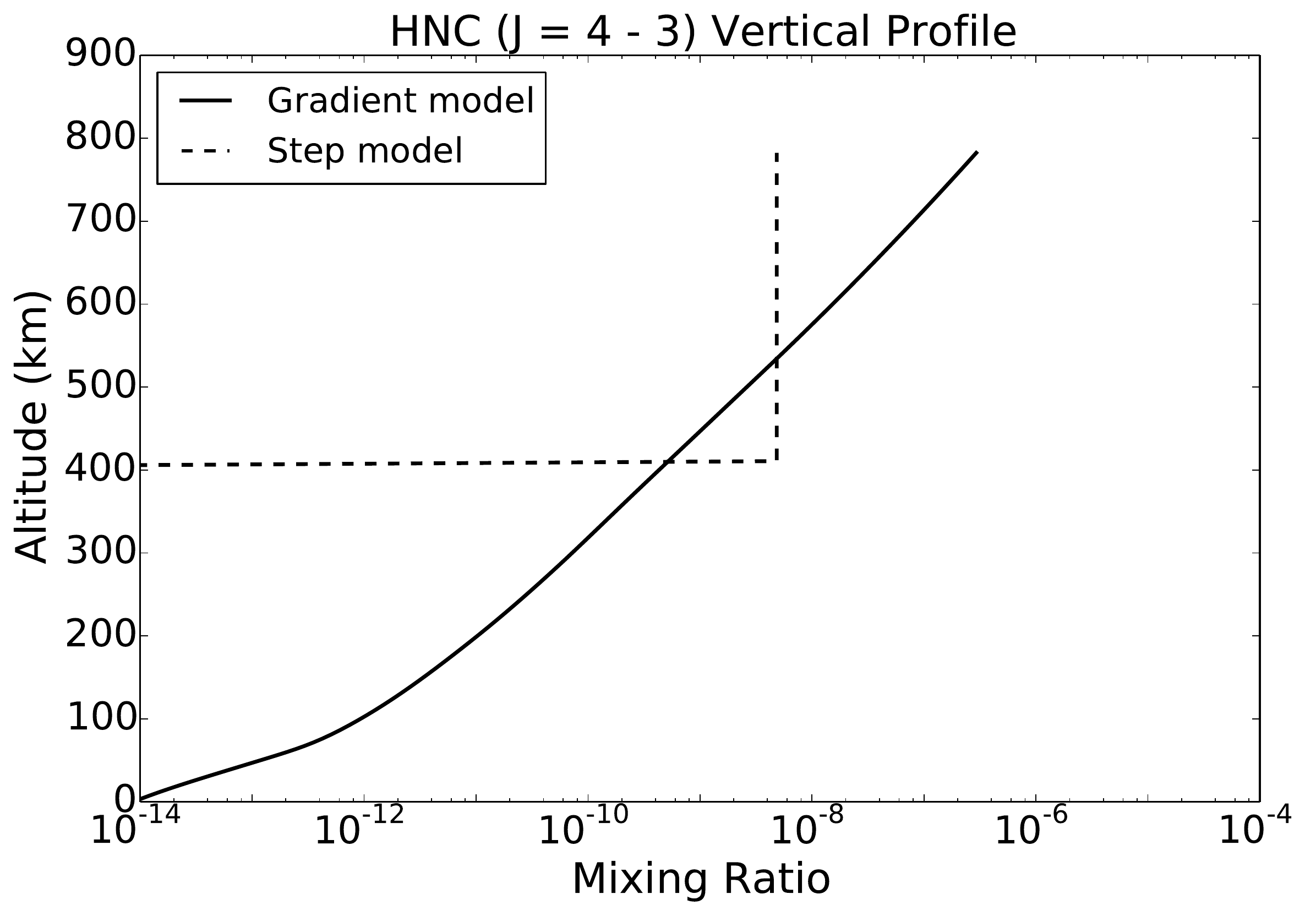}
\hspace{3mm}
\includegraphics[width=0.47\textwidth]{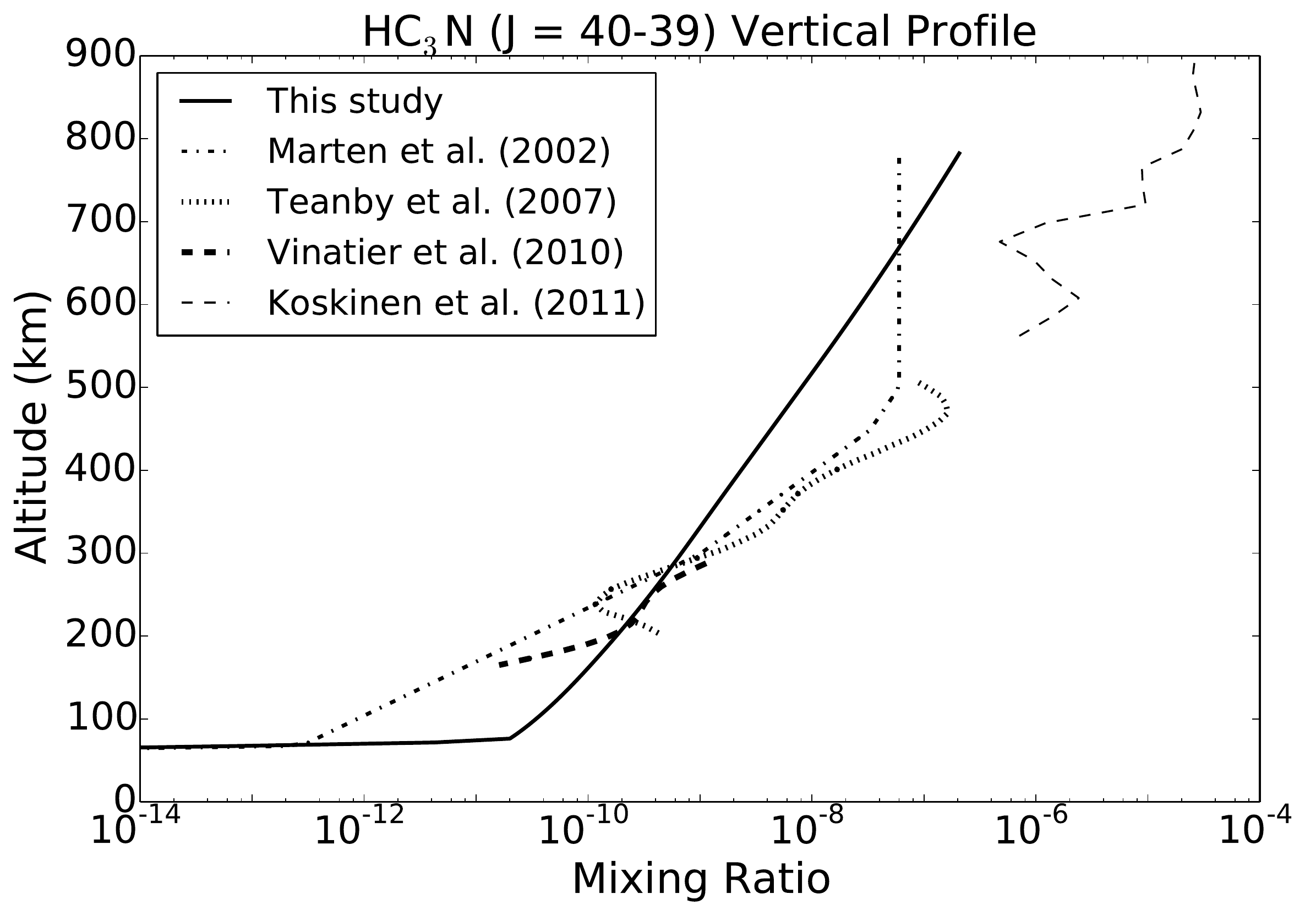}
\caption{Top-left: HNC $J=4-3$ {spatially-integrated} spectrum with best-fitting models overlaid. Top-right: HC$_3$N $J=40-39$ {spatially-integrated} spectrum with best-fitting model overlaid. Bottom-left: Best-fitting HNC VMR profiles. Bottom-right: Best-fitting HC$_3$N VMR profile from this study compared with previous studies.\label{fig:lines}}
\end{figure*}

\begin{figure*}
\centering
\includegraphics[width=0.49\textwidth]{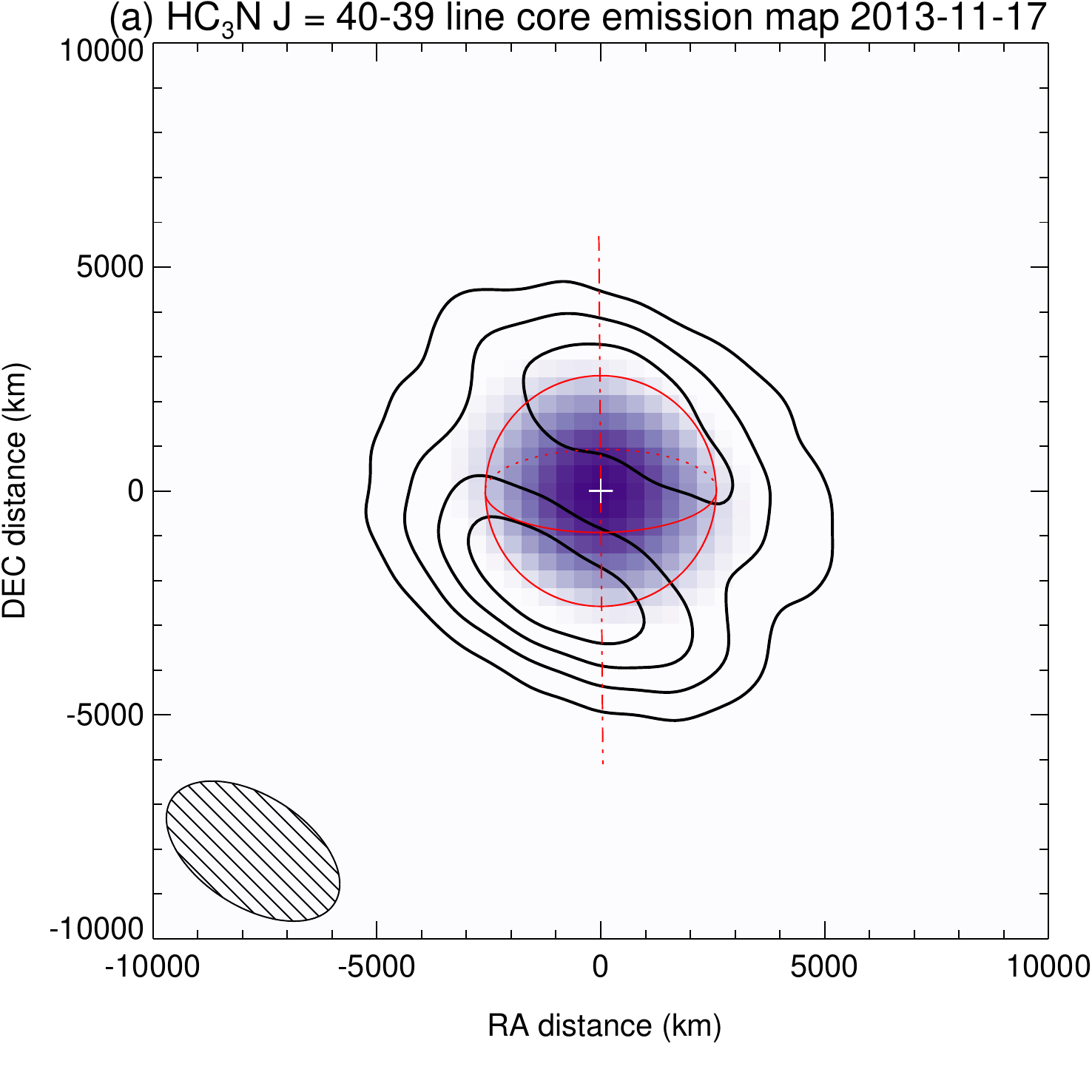}
\includegraphics[width=0.49\textwidth]{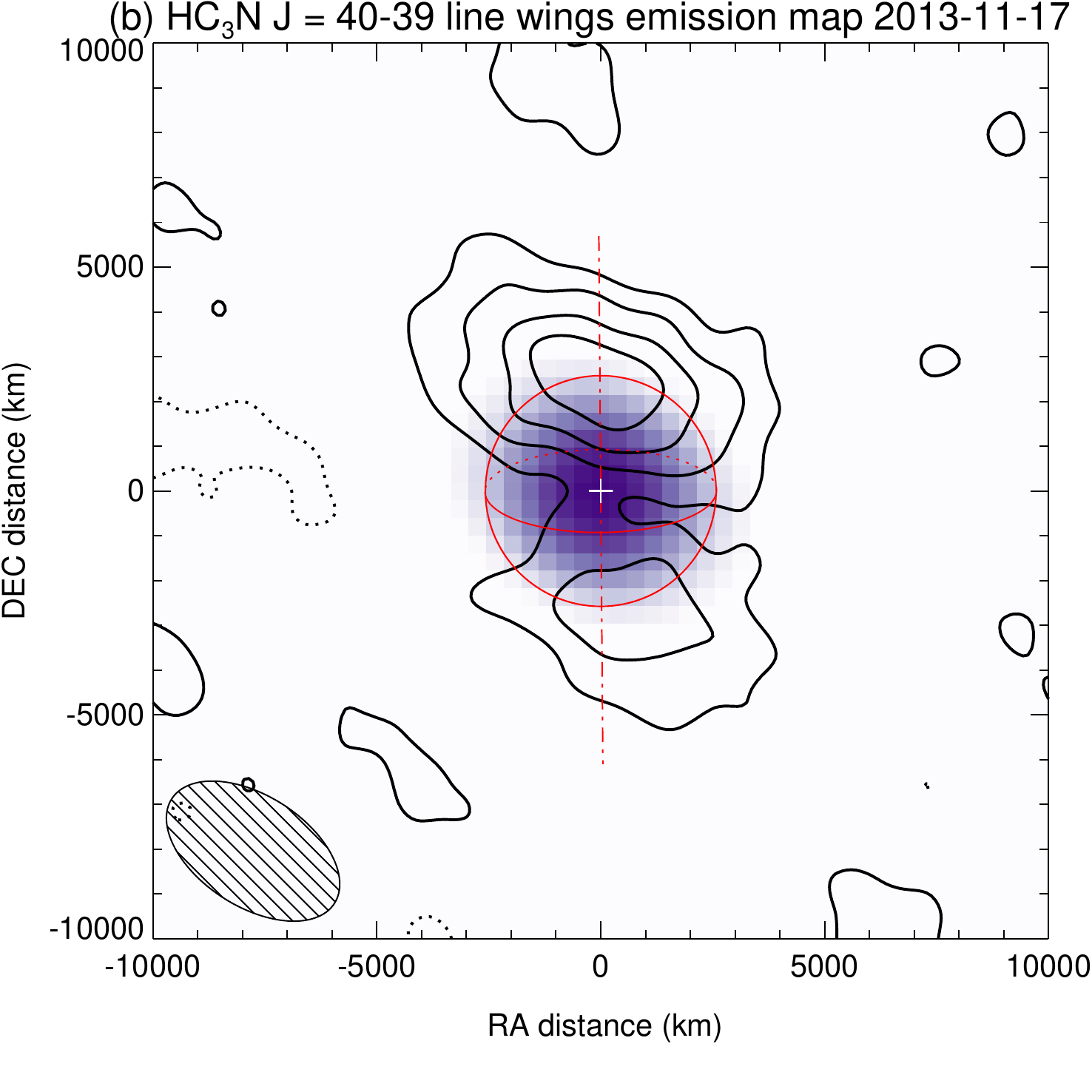}
\caption{(a) Flux contours for the HC$_3$N $J=40-39$ line core (integrated over the range $-0.5$ to 0.5~\kms\ with respect to line center), tracing HC$_3$N at altitudes $z\sim70-600$~km. (b) Flux contours for the HC$_3$N $40-39$ line wings (integration range $\pm$(2.0 to 5.0)~\kms), dominated by low-altitude HC$_3$N ($z\lesssim300$~km). Additional annotations are as in Fig. \ref{fig:maps}. \label{fig:corewings}}
\end{figure*}

\section{Observations}
\label{obs}

Observations of Titan were made on 2013 November 17 in Cycle 1 Early Science mode as part of an ALMA {Director's Discretionary Time} project {to observe comet C/2012 S1 (ISON) \citep[see][for details]{cor14}. To set the flux density scale,} a single 158 second integration of Titan was obtained using the dual-sideband Band 7 receiver. {Despite the short integration time, ALMA's excellent sensitivity resulted in spectra with high signal-to-noise ratio, and we have used these data to obtain the results described in this Letter}. The correlator was configured to cover four basebands, with frequencies in the range 350.34-351.28~GHz and 351.32-352.26~GHz (lower sideband), and 362.18-363.12~GHz and 363.14-364.07~GHz (upper sideband), with 3840 channels per spectral window. The channel spacing was 244~kHz which (after Hanning smoothing by the correlator) produced a spectral resolution of 488~kHz (or $0.40$~\kms\ at 363~GHz).  Weather conditions were excellent, with high atmospheric phase stability, and extremely low precipitable water vapor at zenith (0.52~mm). Quasar {3C\,279} was used for bandpass calibration, and Titan's strong, compact continuum permitted self-calibration of the phases.  The array was configured to track Titan's ephemeris position, updating the coordinates of the phase center in real-time.

{The data were flagged and calibrated by the North American ALMA Science Center using the standard data reduction scripts and protocols provided by the Joint ALMA Observatory. The flux scale was determined by setting the observed continuum level to match} the Butler-JPL-Horizons 2012 flux model for Titan (see ALMA Memo \#594\footnote{https://science.nrao.edu/facilities/alma/aboutALMA/Technology/\\ALMA\_Memo\_Series/alma594/memo594.pdf}). Absolute flux calibration is thus expected to be accurate to within 15\%. Continuum subtraction and imaging were performed using standard routines in the NRAO CASA software version 4.2.0 \citep{mcm07}. Deconvolution (cleaning) of the ALMA point-spread function (PSF) was performed for each spectral channel using the Hogbom algorithm, with natural visibility weighting and a threshold flux level of twice the RMS noise. The image pixel sizes were set to $0.05\times0.05''$. The resulting spatial resolution (FWHM of the Gaussian {restoring beam}) was $0.54''\times0.31''$ (long axis $57.3^{\circ}$ clockwise from celestial north). This resolution element corresponds to $4240\times2440$~km at Titan's geocentric distance of 10.84~AU at the time of observation, compared to Titan's mean radius of 2,576~km.

The images were transformed from celestial coordinates to (projected) spatial distances with respect to the center of Titan, and spectral coordinates were Doppler-shifted to Titan's rest frame using the JPL Horizons Topocentric radial velocity of $-0.91$~\kms.

\section{Results and spectral line modeling}
\label{results}

Emission lines were detected from HNC $J=4-3$ at 362.630~GHz and HC$_3$N $J=40-39$ at 363.785~GHz. Integrated line intensity contour maps are shown in Fig. \ref{fig:maps}. All detected emission is included in the maps, using integration ranges $\pm1.0$~\kms\ with respect to line center for HNC and $\pm5.0$~\kms\ for HC$_3$N. Both species show a clear double-peaked structure, with a separation between the peaks of about $0.6''$ (which projects to 4,700~km on the sky, compared to Titan's diameter of 5,150~km). Whereas the HC$_3$N integrated emission peaks are positioned over Titan's poles, the HNC peaks are each offset by 1500-2000~km in a clockwise direction in the plane of the sky, about the sub-observer point. The signal-to-noise ratios for the HNC and HC$_3$N emission peaks are 7.2 and 11, respectively

Spectra for HNC and HC$_3$N are shown in Fig.~\ref{fig:lines}, {obtained by integrating the observed data cubes over a circular aperture $1.5''$ (12,000~km) in diameter, centered on Titan.} The average RMS noise per channel was {31~mJy} for HNC and {38~mJy} for HC$_3$N. Whereas the HNC line is quite narrow (FWHM~$= 0.76$~\kms), and shows negligible deviation from a Gaussian profile, the HC$_3$N line exhibits broad wings, indicative of collisionally-induced (pressure) broadening, which suggests significant gas abundance in the lower atmosphere/stratosphere.

The {spatially-integrated} HNC and HC$_3$N emission lines were modeled using the {NEMESIS} retrieval code \citep{irw08}, {making use of a recently added line-by-line radiative transfer module.} The atmospheric temperature profile was generated from a combination of Cassini CIRS and HASI measurements \citep{fla05,ful05}, {while the abundances of nitrogen and methane isotopes and aerosols were the same as used by \citet{tea13}}.  Spectral line parameters and partition functions were taken from the JPL catalog\footnote{http://spec.jpl.nasa.gov/}, with frequencies for HNC and HC$_3$N from \citet{oka93} and \citet{tho00} respectively. The Lorentzian broadening half-widths at 300~K were assumed to be 1.290 and 0.1~cm$^{-1}$\,bar$^{-1}$ for HNC and HC$_3$N respectively, with temperature-dependence exponents of 0.69 and 0.75. {Using the model temperature/abundance/aerosol profiles, and making the assumption that these profiles do not vary with latitude and longitude, fluxes were calculated as a function of frequency using a 10-point trapezium-rule integration from the center of the disk to the edge of the model (at an altitude of 782~km).}

{A two-parameter `gradient model' was used for the HNC and HC$_3$N vertical mixing ratio (VMR) profiles, with a variable abundance at reference altitude $z_r = 293$~km, and a variable slope above and below that level, represented as a fraction ($f_H$) of the pressure scale height.} The HC$_3$N was assumed to condense out at altitudes below $z_c \simeq 71$~km, at which saturation vapor pressure is reached for this molecule. Employing a $\chi^2$ minimization between the modeled and observed spectra, the best-fitting abundance at $z_r$ was found to be {$0.62 \pm 0.04$~ppb for HC$_3$N, and $0.063 \pm 0.025$~ppb for HNC, with retrieved fractional scale heights of $f_H = 2.7 \pm 0.2$ and $12.4\pm8.7$, respectively.  The vertical column densities of these best-fitting gradient models were $1.9\times10^{13}$~cm$^{-2}$  for HNC and $2.3\times10^{14}$~cm$^{-2}$ for HC$_3$N.}

As identified by the Herschel HIFI observations of \citet{mor12}, the narrow HNC emission line indicates no significant contribution from this molecule at altitudes $\lesssim400$~km, so we also modeled HNC using a profile with constant abundance above 400~km and zero below. {The resulting `step model' fit is shown in Fig. \ref{fig:lines} (upper left panel; dashed curve), and provides a similarly good fit to the spectrum as the gradient model, although the line wings are slightly weaker.} The best-fitting abundance above 400~km was found to be $4.85\pm0.28$~ppb, consistent with the value of $4.5^{+1.2}_{-1.0}$~ppb obtained by \citet{mor12}. {This model is less optically-thick and has a somewhat smaller vertical HNC column density of $1.2\times10^{13}$~cm$^{-2}$. High signal-to-noise ratio observations of multiple emission lines could help constrain the true HNC distribution.}

The best-fitting VMR profiles derived for HCN and HC$_3$N are shown in Fig. \ref{fig:lines}. For HC$_3$N, profiles from \citet{mar02}, \citet{tea07}, \citet{vin10} and \citet{kos11} are plotted for comparison. Our HC$_3$N VMR profile is in reasonable agreement with these previous studies, although we identify significantly more HC$_3$N in the lower stratosphere ($z\sim70$-200~km) than \citet{mar02}. In addition to variations in the degree of spatial averaging between the different studies, seasonal and latitudinal variations in the HC$_3$N abundance distribution can plausibly be responsible for the differences between the observed profiles.

\section{Discussion}

Peaks in the integrated HC$_3$N emission over Titan's poles are interpreted as resulting from enhancements in the HC$_3$N abundance that arise as a consequence of global atmospheric circulation. Cassini CIRS observations identified the presence of enhancements in hydrocarbon and nitrile abundances over Titan's northern (winter) pole during the period 2004-2007 \citep{tea08,tea10,cou10,vin10}. These abundance patterns were found to be consistent with the presence of a global south-north circulatory flow (similar to Earth's Hadley Cells), that carries high-altitude photochemical products towards the winter pole. Following the August 2009 equinox, the circulation cell began to undergo reversal (from northern to southern subsidence), which manifested as changes in the molecular distribution patterns observed by CIRS between 2010 and 2011 \citep{tea12,cou13}.  

The appearance of both a northern and southern polar peak in the HC$_3$N distribution measured by ALMA in November 2013 may be interpreted in two ways: 

(1) A recent reversal of the main high-altitude circulation cell could be responsible for transporting HC$_3$N from its production site in the upper atmosphere towards the south pole, where it becomes concentrated as a result of subsidence. Chemical models have identified a lifetime for HC$_3$N of $\sim1.6$~yr \citep{kra09}. This calculation is based on conditions near the equator; chemical lifetimes at the poles may be longer due to the reduced Solar insolation, so the northern HC$_3$N peak can plausibly be a remnant of the HC$_3$N transported to that region prior to the recent reversal in the global circulation. If so, a steady decay in the northern HC$_3$N abundance would be expected over the next few years, measurement of which could provide a quantitative test of HC$_3$N's chemical lifetime near the pole.

(2) Alternatively, the presence of two circulation cells (rising near the equator and moving in opposite directions) may be possible during the seasonal transition period \citep[\eg][]{ran05}. Such flows could carry photochemical products (including HC$_3$N) from the upper atmosphere towards both poles simultaneously. Observations of variations in the HC$_3$N distribution over the course of Titan's seasonal cycle (for example, covering the completion of the transition from northern spring to the summer solstice in 2017) could help determine the origin of the observed structures and provide new insight into Titan's global circulation patterns.

Similar to HC$_3$N, HNC also shows a clear double-peaked structure. However, in contrast to HC$_3$N, the observed HNC peaks are not aligned with Titan's polar axis, but are offset clockwise by $40\pm5^{\circ}$ in the plane of the sky. These offsets imply longitudinal and latitudinal variations in the HNC distribution/excitation that are not readily explainable in the current paradigm of chemical and circulation models, which typically possess longitudinal symmetry. It seems contrived to postulate enhanced HNC production and/or reduced destruction rates at specific latitudes and longitudes, or the presence of a peculiar HNC circulation system that concentrates this molecule at the observed positions. Titan's super-rotating zonal winds (with speeds $\approx60$~m\,s$^{-1}$ at 450~km altitude; \citealt{mor05}), should result in longitudinal mixing of the atmosphere on a timescale of about 45~hr, which renders diurnal chemical differentiation an unlikely explanation for the observed structures, given Titan's 382~hr day length.   Variations in the atmospheric temperature profile as a function of longitude could plausibly introduce corresponding changes in the HNC excitation and thus, the spectral line intensity, but the origin of such asymmetrical variations across Titan's disk presents a challenge for our current understanding. 

Variability in the locations of peak HC$_3$N emission with altitude may assist in the interpretation of the mysterious HNC distribution.  Due to the variation in line pressure-broadening as a function of altitude, maps can be constructed of the HC$_3$N emission as a function of altitude. As demonstrated by \citet{mar02} --- see their Fig. 1 --- the HC$_3$N $J=26-25$ line wings (at frequencies $>1.6$~MHz, or $>2$~\kms\ from line-center), trace gas at low altitudes ($z\lesssim300$~km), whereas the line core traces gas at $z\sim80-600$~km. In Fig.~\ref{fig:corewings}a, we present maps of the HC$_3$N line flux integrated over (a) just the line core ($-0.5$ to 0.5~\kms) and (b) just the line wings (2.0 to 5.0~\kms\ either side of the line center).  Given that the low-altitude HC$_3$N distribution (Fig. \ref{fig:corewings}b), as well as the total integrated HC$_3$N flux (Fig. \ref{fig:maps}) are relatively symmetric about Titan's polar axis, the asymmetry in the HC$_3$N distribution in Fig. \ref{fig:corewings}a (in which a line through the HC$_3$N peaks is tilted clockwise by about $25\pm5^{\circ}$ with respect to the poles) implies that the emission from high-altitude HC$_3$N (at $z\gtrsim300$~km) must be offset from the poles in a similar way to the HNC. Thus, it can be concluded that the asymmetrical HNC distribution results from a process that becomes important only in the mesosphere and above. The fact that the emission maps for these two chemically-distinct species are affected in a similar way implies that the observed structures are more likely the result of a physical phenomenon such as temperature or atmospheric dynamics, rather than a consequence of atmospheric chemistry.

\section{Conclusion}

We report the first detections of molecular emission from Titan's atmosphere using ALMA and present spatially-resolved maps of emission from HNC and HC$_3$N. These maps show remarkable double-lobed structures for both molecules, with clear concentrations of HC$_3$N {at low altitude ($z\lesssim300$~km)} over Titan's north and south poles, whereas the HNC and high-altitude HC$_3$N emission are offset from the poles by 20-45$^{\circ}$ in a clockwise direction. Disk-averaged atmospheric emission models that take into account the shapes of the resolved spectral line profiles confirm the result of \citet{mor12} that HNC is predominantly confined to altitudes $\gtrsim400$~km. Similar to previous radio and infrared studies, HC$_3$N is shown to be abundant at altitudes 70-500~km, with a VMR profile that increases with altitude up to {at least} 500~km. These maps provide the most striking example to-date of longitudinally asymmetric molecular emission in Titan's mesosphere. The implied longitudinal anisotropies present a challenge for current understanding of Titan's atmospheric structure and circulation, and indicate that the mesosphere may be more longitudinally-variable than previously thought.

This study demonstrates ALMA's unique ability to perform spatially and spectrally-resolved measurements of molecular emission from the atmospheres of Solar System bodies. {Using less than three minutes of observational data, our results also highlight the extremely high sensitivity of ALMA, even in Early Science mode}. Future interferometric observations of Titan's atmosphere at higher angular resolution and with improved sensitivity will provide new insights into its structure, composition and dynamics, which are expected to facilitate our understanding of the complex interplay between photochemistry and circulation {in planetary atmospheres}, with applications within our Solar System and beyond.

\acknowledgments
This research was supported by NASA's Planetary Atmospheres, Planetary Astronomy and Astrobiology Programs, by The Leverhulme Trust and the UK Science and Technology Facilities Council. The article makes use of the following ALMA data: ADS/JAO.ALMA\#2012.A.00033.S. ALMA is a partnership of ESO (representing its member states), NSF (USA) and NINS (Japan), together with NRC (Canada) and NSC and ASIAA (Taiwan), in cooperation with the Republic of Chile. The Joint ALMA Observatory is operated by ESO, AUI/NRAO and NAOJ. The National Radio Astronomy Observatory is a facility of the National Science Foundation operated under cooperative agreement by Associated Universities, Inc.

\end{document}